\documentclass[conference]{IEEEtran}
\usepackage{blindtext, graphicx}
\ifCLASSINFOpdf
\else
\fi
\usepackage{multicol}

\begin{document}
\title{CDMA for Underwater Acoustic Communication}
\author{\IEEEauthorblockN{Lokesh Bommisetty, Samskruthi Gaddam, Nomula Prakash Reddy, Shaik Basharath and Bharath Are }
\IEEEauthorblockA{National Institute of Technology Goa, Goa, India}
Email: {(lokesh.jun12, samskruthichandra,prakash.nitgoa24,basharath.nitg,bharath.are17)}@gmail.com,}

\maketitle
\begin{abstract}

This paper provides an overview of how and why acoustic waves are important for underwater communication when compared with radio and optical waves, underwater acoustic channel, it's characteristics and model. Also about the underwater sensor architecture, challenges faced by sensor nodes in the network and Media Acess Control(MAC) protocol solutions for them. This paper also describes about the various communication approaches including channel coding in underwater and their performance analysis for different approaches used. Through these approaches, multiple underwater vehicles can communicate among themselves as well as with the console with minimal Bit Error Rate.\\

%

\end{abstract}
\section{INTRODUCTION}

Underwater communication is a technique for transmitting and receiving the information signals below water. The difficulties faced in this are due to time variations of the channel, multi-path propagation, low bandwidth availability and strong signal attenuation,especially over long ranges \cite{16}. \\
Unlike for terrestrial communications Electromagnetic (EM) waves are not very favourable for carrying out underwater communication. Attenuation of radio waves in water gets increased with increase in both conductivity and frequency as in the case of any other medium. The possible solution can be rising an antenna of considerable length above the sea level and then to use ordinary low frequency radio waves. In fact, radio waves propagate long distances through conductive sea water only at extra low frequencies of 30 - 300 Hz, which require large antennae and high transmission power.\\
Very Low Frequency radio waves of 3–30 kHz can penetrate through seawater to a depth of approximately 20 metres. Hence a submarine at shallow depth can use these frequencies \cite{13},\cite{15}. Because of the narrow bandwidth of this band, VLF radio signals cannot carry audio (voice) signals, and only can transmit text messages at a slow data rate. Unlike radio waves Optical signals do not suffer from such high attenuation but are affected by scattering.\\
 With an idea of avoiding such extra resources acoustic waves can be used which provides the most obvious medium to enable underwater communications. Acoustic propagation is best supported at low frequencies, and the bandwidth available for communication is extremely limited.\\ 
   In general the modulation methods that are developed for radio communications can be very well adapted for underwater acoustic communications (UAC) \cite{11}.

High-speed underwater acoustic channel communication is challenging due to extended multi-path propagation, limited bandwidth,, severe fading, rapid time-variation, refractive properties of the medium and large Doppler shifts. In order to make the well use of the limited bandwidth available, Code Division Multiple Acess(CDMA) technique can be employed in which bandwidth can be exploited at its maximum \cite{12}.

\section{Underwater Acoustioc Channel}
In any communication system,Channel is the very important block to be analysed because all the other blocks of the system should be modelled in accordance with channel. Channel characteristics have great effect on of communication. Also, the same channel shows different characteristics for different signals such as radio waves, optical waves, acoustic waves passing through it. Here, in this section, underwater acoustic channel is analyzed.
\subsection{Characteristics of Underwater Acoustic Channel}  
Underwater acoustic channels is considered to be one of the most complicated media of communication in use\cite{1}. Acoustic communication shows better performance at low frequencies of operation. Also the bandwidth available for acoustic communication is extremely limited. The following characteristics play prominent role in deciding the other communication parameters of the system.
\subsubsection{Attenuation and Noise}
\begin{figure}
    \centering
    \includegraphics[width=\linewidth]{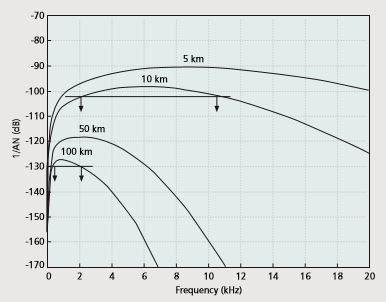}
    \caption{SNR with respect to frequency and distance in acoustic channel}
    \label{fig_sim}
\end{figure}

The path loss,absorption and diffusion of acoustic energy depends on the transmitted acoustic signal frequency and the distance traversed by the signal. In addition to these losses, signal experiences spreading loss which increases with distance traversed. The path loss is given by
\begin{equation}
    \centering
A(l, f)=(l/l_{r})^{k}a(f)^{l-l_{r}}
\end{equation}
where signal frequency is denoted by $f$ and distance traversed by signal by $l$ which taken in reference to some reference distance  $l_r$. The path loss exponent k which has values usually between 1 and 2, models the spreading loss \cite{2}. The SNR in acoustic channel dependence on both distance and frequency is shown in Fig. \ref{fig_sim}.\\

Noise in underwater acoustic channel comprises of ambient noise and site-specific noise. Ambient noise is a Gaussian noise which will be present everywhere and it comes from the sources like turbulence, breaking waves, rain, and distant shipping. Site-specific noise will be present at only certain cites and it is non-gaussian.
\subsubsection{Multipath}
The transmitted acoustic signal gets reflected at surface or bottom of the water body or at any objects in water,refracts due to the spatial variability of sound speed  and hence results in the multipath propagation. Multipapth propagation results in fast fading,frequency selective fading of the signal which can be overcome by different techniques such as Equalization Technique, Spread Spectrum Technique, Diversity Technique and Array Technique \cite{2}.
   The physical geometry of the channel and properties like reflection and refraction influences the impulse response of the acoustic channel, which determine the number of  propagation paths those are significant, and their relative signal strengths and path delays. Broadly speaking infinitely many echos can occur in the channel forming infinite paths, but subjecting it to the signal strengths the number of paths are limited to a finite value. 

To analyse channel in mathematical model, consider the $p^{th}$ of propagation path has a length of $l_p$, path delay as
\begin{equation}
\centering
  \tau_p=l_p/c, 
\end{equation}
where $c$ is the speed of sound in shallow water, considering to be a constant.Now the frequency response of the $p^{th}$ path can be represented as 
\begin{equation}
\centering
    H_{p}(f)={\Gamma_{p}\over \sqrt{A(l_{p}, f)}} ,
\end{equation}

where numerator is the cumulative reflection coefficient and $A(l_p,f)$ is the propagation loss in $p^{th}$ path.

 Hence, each path of an acoustic channel acts as a low-pass filter, which contributes to the overall impulse response,
 \begin{equation}
 \centering
     h(t)=\sum\limits_{p}h_{p}(t-\tau_{p}),
 \end{equation}
 
where $h_p(t)$ is the inverse Fourier transform of $H_p(f)$.

\subsubsection{Time Variability and Doppler Shift}
Channel's time variability majorly due to two sources:inherent changes in propagation medium and changes due to motion in transmitter or receiver. Inherent changes vary from short timescales that affect the instantaneous signal level to very long timescales. The inherent changes of short timescales are prominently induced by surface water waves.\\
Motion of transmitter or receiver additionally contributes Doppler effect on the transmitted signal. Frequency spreading and shifting of information signal occurs due to the Doppler effect on it. As the speed of sound waves(acoustic waves) is very less compared to the velocity of electromagnetic signal, the effect of Doppler effect will be magnificent in case of acoustic signal \cite{3},\cite{4}.\\
Proper channelling model can overcome the problems resulting from these errors.
\subsection{Model of Underwater Acoustic Channel}
Underwater acoustic channel is a slow time varying, coherent and a multipath channel. Within the coherence time, the channel can be approximated as a coherent multipath channel which means that the channel does not suffer from inter symbol interference in a particular path but it has multipath effect.\\
If the transmitted signal is $s(t)$ then the received signal is given as
\begin{equation}
\centering
r(t)=\sum\limits_{k=1}^{L}a_{k}(t)s(t-\tau_{k}(t))+n(t),
\end{equation}

where $k$ is the propagation path among $L$ paths, $a_k$ and $\tau_k$ are the attenuation factor and time delay in $k^{th}$ path and $n(t)$ is the additive noise.

\begin{figure}
    \includegraphics[width=\linewidth]{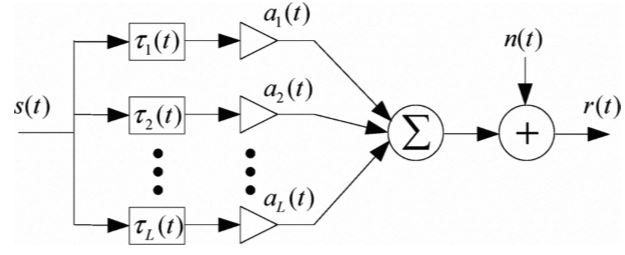}
    \caption{Model of acoustic channel with additive noise}
    \end{figure}

\section{ARCHITECTURE OF UNDERWATER WIRELESS SENSOR NODES}

Underwater sensor network’s architecture is not fixed, provisional and wavering one. Four different types of sensor nodes are presently used in the architecture. A large number of variety of sensors subject to check water pressure, change in salinity of water etc., are positioned at lowermost layer at the seafloor \cite{5}. They communicate with the surrounding nodes and also collect data from the sensors at seabed through the acoustic modems. This layer is supplied power with the batteries and they tend to operate for longer periods of time because they operate periodically. These nodes are equipped with distributed localisation algorithms through which they are able to determine their locations and communicate.

The top layer contains the control section of the architecture, control nodes with proper connections. This node is setup at off-shore platform with power or on-shore in which nodes have large storage capacity to access and buffer the data with upright electrical power. These nodes are connected directly to acoustic modems with wires through which they establish communication with sensor nodes. The architecture of sensor nodes is shown in Fig.3.

\begin{figure}
\includegraphics[width=\linewidth]{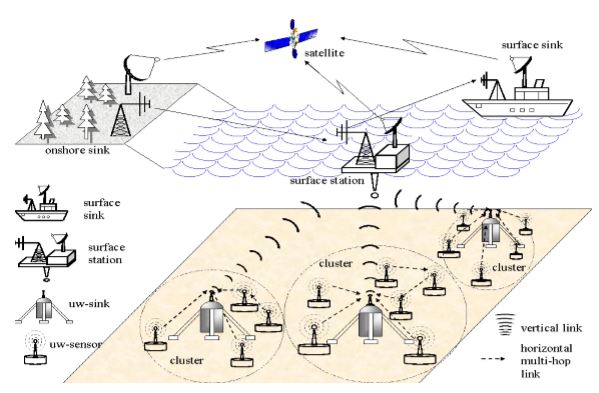}
\caption{Architecture of underwater sensors}
\end{figure}
There are nodes with large networks called super nodes which advance over the area. It’s access data in well-organized manner and work systematically and effectively at high speeds to relay the information to the base station. These nodes help attach the usual nodes to node buoys, mark of the anchored float to provide high speed communication with base station. These nodes are especially needed for higher network connectivity and establish as many data collections as possible \cite{21}.

The underwater rough and severe environment greatly affect the underwater sensor network. After a period of time some nodes may be lost or get damaged by fishing vessels, underwater life, sinking of sensor nodes into seabed. 

\subsection{Challenges in Underwater Sensor Networks:}
\begin{itemize}
  \item  To acquire strong and effectively suited and nanosensors and secure less expensive devices.
\item To employ the mechanisms which work against corrosion and cause pollution during the periodic cleaning of the underwater devices.
\item To ensure the safety of physical, Chemical and biological factors which influence the sub adjacent marine usage system.
\item Sensors should be stable and effective enough to withstand in high degrees of temperature.
\item The stern and strict restriction of the availability of bandwidth.
\item To withstand the short-term bit errors and loss of data during transmission.
\item The worse conditions of underwater channel by the multipath transmission and fading phenomena.
\item The higher magnitudes of propagation delays in underwater acoustic channel when compared RF in terrestrial communication networks.
\item Proper localization of sensors to ensure exchange information at high speeds effectively.
\item Reliability of data delivery from deep level of underwater to surface level.
\item The lifetime of the battery power and replacement costs as the underwater batteries are not chargeable and greatly affected by the corrosion and pollution in water.
\item The need for high amount of power for communication apart from maintenance of sensors.
\item The high cost of the hardware protection equipment required for protection of necessary devices of network.
\end{itemize}

\subsection{Limitations of Generalized Application of Underwater Sensor Network:}
\begin{itemize}
 \item  Collection of information or data samples from aquatic environment like river, lake and oceans.
\item Exchange of data from one node to another through nodal gateways.
\item The administration and prior survey of the area of established sensor network.
\item Immediate recovery of any issues within system and targeting intrusion detected by system.
\item Investigation of aquatic environment which is not possible for humans to explore and monitoring of chemical, biological pollutants.
\item Detection of minerals, oil fields and natural herbs within deep water for medicine.
\item To prevent dry land from the disasters like Tsunami, sea quake, sudden high tide and low tide etc.,
\end{itemize}

\section{Communication Approaches}
\subsection{Modulation Techniques}
Major challenge of acoustic communication is limited bandwidth. In FSK modulation technique  frequencies corresponding to digital zero and one bits are chosen irrespective of channel variations. But in FSK to  avoid overlapping of  frequencies used, place for guard bands is left in between them. As acoustic channel has limited bandwidth FSK is not entertained for acoustic communication. Also acoustic channels has the problem of Doppler spread hence OFDM cannot be employed. With the help of QAM and PSK data rates of uw channel can be improved but they add the issue of inter symbol interference\cite{6}. CDMA and MIMO are the emerging modulation schemes in uw communication. CDMA helps to differentiate signals sent by various devices, and reduces the effects the multipath propagation. CDMA helps in achieving maximum throughput and aims for secure transmission of messages\cite{7}.

 \subsection{UNDERWATER MAC PROTOCOL} 

Sensors and underwater vehicles are assumed as nodes or terminals of a wireless network that share information. Sensors are powered with limited battery power. Every time a packet of data is sent or received sensors consume some of their battery power. There is power wastage in sensor nodes due to collision, overhearing or idle listening. Media Access Control protocol provides solution to the above mentioned issues.
\begin{figure}
\includegraphics[width=\linewidth]{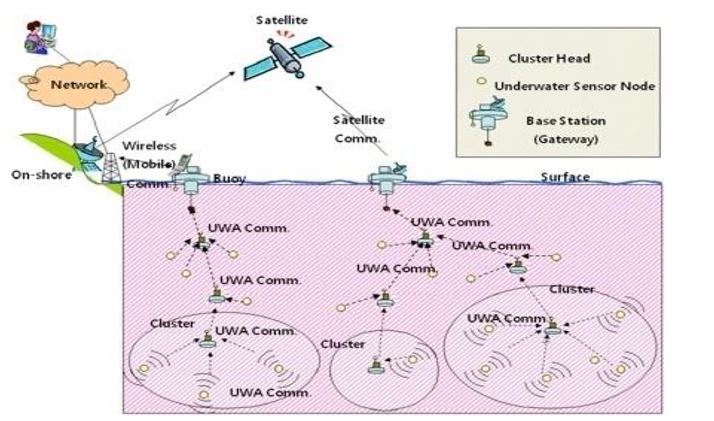}
\caption{Underwater Acoustic Communication.}
\end{figure}
Medium Access Control provides channel access for the nodes. MAC protocol can be broadly classified as contention based or contention free. Some of the contention free MAC protocols are FDMA, TDMA, CDMA etc.,\\
\subsubsection{FDMA}
 In Frequency Division Multiple Access all the available bandwidth is split into smaller frequency bands where each node is assigned with certain bands for all the time. In the case of underwater communication available bandwidth is very limited i.e., from 3-30khz. If this bandwidth is to be shared among several sensor nodes of the network each node may get a very narrow bandwidth. It may be subjected to complete fading also increasing the risk of interference due to very narrow bands. In Time Division Multiple Access time is divided into slots where each node is assigned a particular time slot. When a node gets it’s turn to it may use all the available bandwidth of the channel\cite{10}.
\subsubsection{TDMA}
In Time Division Multiple Access time is divided into slots where each node is assigned a particular time slot. When a node gets it’s turn to it may use all the available bandwidth of the channel. MAC protocol says each node to have periodic listen and sleep times to prolong life of sensors.

Nodes can transmit or receive packets of data in their listen times, thereby avoiding issue of idle listening. In sleep times, sensors are not active saving power. Every node should be aware of listen and wake up times of other nodes in the network for sender node not to transmit packet during sleep time of receiving node to avoid resending. To resolve this problem MAC protocol provides with schedule maintenance. Each node  has a table of its own listen and sleep time along with schedule of other nodes. Nodes have freedom to  chose their own schedules by looking into free slots left by other nodes. Every node waits for some time to listen to the incoming schedules given by other nodes. If it does not receive any such schedules it creates a schedule for itself and broadcasts it other nodes.

Nodes broadcast their schedule with SYNC packet with its node id  along with time after which it is going to sleep. In case a particular node want to create its own schedule but received a schedule from other nodes it will fill into it’s table the schedule time of other nodes and also broadcast it’s schedule using SYNC packet. The node which starts setting schedule first is called synchronizer and others followers. Schedules are broadcasted periodically. When the node is active listen time  is divided into two parts. One is SYNC packet slot where they send information about their schedule and the other data packet slot. The listen time of receiver is further divided into tiny time slots. Sender will randomly pick some tiny time slot of receiver’s SYNC slot  and wait till the end of that period to check if the channel is free. If channel is free it will send SYNC packet to the receiver. Further to send information sender will randomly pick tiny time slot of data slot and wait till the end of the slot to check if channel is free. If it is free it will first  send a RTS-request to send –packet to receiving node. RTS packet will have details of sender node, receiver node and the duration for which they will communicate. Receiving node will acknowledge the RTS packet by sending CTS-clear to send-packet to the sender. When other nodes in the range of  sender hear of RTS packet sent by sender to a receiver they will mute themselves for the duration mentioned in RTS to save energy and thereby idle listening can be avoided. Also nodes in the range of either sender or receiver need to sleep after receiving RTS or CTS packet  to avoid interference. Nodes other than sender, receiver and nodes in the range of sender or receiver, can transmit or receive  to or from any other nodes in the network except for the above mentioned nodes to interference.
\subsubsection{DS-CDMA}
 DS-CDMA is a spread spectrum technique in which usually the bandwidth the code generated is far wider than the message signal. In DS-CDMA MAC, sender nodes in the network can randomly access the channel to send Extended Header(EH). EH is a short header containing the details about sender and receiver nodes and parameters used to construct spreading code used by sender node to spread the message signal. Once the EH packet is sent successfully, sender node then transmits the data packet. With the help of parameters sent by sender, receiving node will construct spreading sequence to multiply it with received signal to get back the original signal transmitted by sender node. Once the receiver node receives the data packet correctly it then sends an acknowledgement to the sender node in the form of acknowledgement packet. ACK packets helps to resolve the issues of resending of packets thereby saving power which may be lost due to retransmissions and also decreases propagation delays by avoiding retransmissions \cite{8}.
MAC protocol hence solves the problems of excess energy consumption for resending the packets, avoids collision and increases the number of successful transmissions \cite{9}.

CDMA is an example of multiple access method, used by various radio communication technologies, where several transmitters(users) can send and receive information simultaneously over a single communication channel making the effective use of the channel.

There are sequences of codes known as PN Codes which are used for spreading the messages sent from users. PN codes for different users are different and mutually  orthogonal.

At the receiving end, the part of signal that has the same PN sequence(absolutely) as that of the user is adopted and received correctly. And the part of the signal that is connected with other PN sequences is discarded as noise because only matching PN codes are correlative.

There are three commonly used spreading techniques. They are DS (Direct Sequence), FH (Frequency Hopping) and TH (Time Hopping) spread spectrum \cite{5}. In the Direct Sequence-CDMA system, with the help of the wide-band PN code original signal is modulated linearly. It is the widely used method for spreading the frequency spectrum.

\subsection{Channel Coding and Decoding}

In underwater acoustic channel, due to the physical properties of channel, salinity of water, and other factors the transmitted data may be affected to noise and can get flipped at the receiver end. This is called an occurrence of bit error. The performance of a communication link is measured in terms of Bit Error Rate (BER) which is the probability of a bit being received is in error. The mentioned properties of channel adds noise to the transmitting signal. Acoustic channel results in two types of noises predominantly. They are Random Error and Burst Error.Channel coding is applied and extra bits are introduced to the existing data bits to detect,rectify or to discard the error occurrences due to channel behaviour.
\subsubsection{Random Errors-RS code}
Random error occurs in random channels such as Additive white Gaussian noise channel where the noise occurs at random time and amplitude. All the errors that occur are randomly independent to each other. To solve this problem Reed-Solomon code is used widely for acoustic channel.

RS code is the origin BCH code which come under the category of Error Correction codes. The raw analog data is quantized and source coded primarily. During this process each sample will be coded into a symbol. Let us consider each symbol consists of $s$ bits.Generally, RS code is denoted by ${RS[n,k]}$. $n$ is the codeword length that is the number of symbols in each codeword after channel coding. $k$ denotes the number of data symbols in each codeword. The remaining $2t=n-k$ symbols are the Forward error correction symbols which are determined by the Reed-Solomon algorithm. Codeword length is decided on the basis of number of bits in each symbol as
 \begin{equation}
 n=2^{s}-1.
 \end{equation}This code can detect and correct at most $t$ error symbols or it can erase at most $2t$ error symbols.

The RS code completely based on the extended Galois fields, which are used to generate GF polynomial. RS codewords are generated on the basis of three polynomials namely GF polynomial, Generator polynomial and Encoding polynomial. Encoding polynomial which is generated on the basis of GF field elements and generator polynomial encodes the incoming data bits to a length of $n$ out of which the first $k$ symbols are the data symbols.The parameters of code for a set of values are listed in Table \ref{tab:my_label}. The look up tables for each galois field are maintained at transmitter as well as receiver, with whose knowledge symbol errors can be detected and rectified or erased sometimes.

\begin{table}[]
    \centering
    \caption{Parameters of RS code}
    \begin{tabular}{|c|c|c|}
    \hline
         $k$& $t$&$g(x)$ \\
         \hline
         11&1&23\\\hline
         7&2&721\\\hline
         5&3&2467\\\hline
         1&7&77777 \\
         \hline
         
    \end{tabular}
    \label{tab:my_label}
\end{table}
\subsubsection{Burst Errors-Interleaving}
Burst errors are the errors that occur at a particular instant of time affecting the bits those are together. These errors are generally caused by impulse noise and fading of the channel. As all the data bits those are affected by noise are at same place even RS code cannot rectify them since RS code can detect a maximum of $t$ symbols for every $n$ symbols.

So the immediate logic to be applied is to distribute the error bits among the whole data bits so that they are well separated from each other which makes RS code capable of correcting the error bits. So the data is to be shuffled well for the error bits to get distributed. Interleaving technique is used to achieve this purpose. There are different types of interleaving techniques such as matrix interleaving,convolution interleaving,block interleaving etc., Fig.5, 6 shows how interleaving can distribute the error symbols in an English phrase transmitted.\\
\begin{figure}[h]
\includegraphics[width=\linewidth]{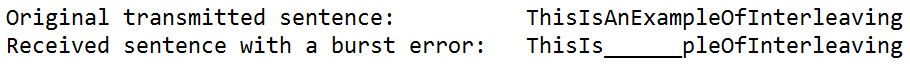}
\caption{Transmission without interleaving}
\end{figure}
\begin{figure}[h]
\includegraphics[width=\linewidth]{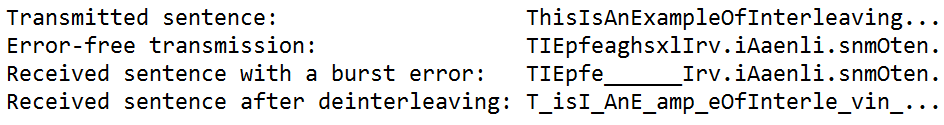}
\caption{Transmission with interleaving}
\end{figure}
\section{Performance Analysis}
Space-Division Multiple Access (SDMA) for CDMA Underwater Acoustic Communication (UAC) employs both Time Reversal Mirror (TRM) and Active Average Sound Intensity (AASI) to effectively improve the capacity of CDMA system and achieve low bit error communication. TRM mainly focus on desired user among multi-user by making use of the physical characteristics of underwater acoustic channel. Whereas the AASI detector can evaluate the exact position of all users with same frequency at the same time by vector combination to get directional communication [3].
Mainly here CDMA In-phase components are computed by varying the Bit Error Rate (BER) and Signal to Noise Ratio (SNR) factors. And the performance of the Multiuser System by SDMA-CDMA is compared with normal CDMA and TRM-CDMA methods. The performance of system is improved because the mutual correlation among the various users is greatly decreased due to the involvement of AASI. Moreover, AASI is easy to implement and very low amount of power is required for UAC.

CDMA based Medium Access Control (MAC) for underground sensor network focus on achieving high network throughput, lower energy consumption and lower delay for accessing channel. These factors may not much be concerning in terrestrial communication but they play great role for proper underwater communication. In this code length and optimal transmit power are set by closed loop distributed algorithm  \cite{17}. 
In CDMA based MAC, the minimum amount energy required per bit for various code lengths is computed Moreover average amount of packet delay, normalized used energy, normalized successfully received packets for different number of sensors is analysed.

Quaternary Phase Shift Keying (QPSK)-CDMA Underwater Acoustic System helps the multiple vehicle to communicate each other without interference at the same time. The synchronized channel of underwater acoustic channel with involvement of the Radio frequency modulation with rake receiver is designed \cite{18}.
In QPSK-CDMA it mainly concentrates on the decreasing BER by finding the optimised SNR with AWGN as noise for number of user in CDMA system.

Block Interleaver – It is able to rearrange the various elements of the input without any repetitions and without discarding any one of them \cite{19}.
Convolutional Interleaver – It consists of a pack of shift registers each with fixed amount of delay and buffer for memory purpose. In this each time the new symbol arrives it is taken into the shift register whereas the present symbol present in the register is sent out as a part of the output vector. It requires both present and previous symbols for analysis of each and every code. \\   
Matrix Interleaver – It consists of a matrix block for internal computation on the code patterns of the symbols and the matrix is filled with input symbols row by row whereas the matrix block sends the output of calculated data column by column and order of the matrix is given by the number of rows of input by number of outputs it delivers through output port .
The performance of system is measured by the amount of loss of bits and the amount of bit error for various by changing the each interleaver along with normal CDMA processing of codes and it is noted that Convolutional interleaver achieves low bit error and low amount of loss of bits \cite{20}.

\section{Conclusion}
This paper illustrates different techniques that can effectively implement underwater acoustic communication.Limitations of communication signals in underwater medium and the characteristics of channel are discussed briefly in the paper.Acoustic communication based on MAC protocol and DS-CDMA is explained in detail.As the acoustic channel is very complex a strong channel encoding scheme should be implemented to overcome the errors introduced by the channel, which are also explained in this paper.

The design of rake receiver to take advantage of multipath propagation and synchronization of PN sequence at the receiver end is also very important in improvising the output SNR and to reduce Bit Error Rate(BER) considerably.These issues are not taken into consideration in this paper, which are worth doing research in future.

%
%
%
%
%
%
%
%
%
%





\end{document}